\documentclass[10pt,twoside]{Lowx2011}
\usepackage{epsf,amsmath}
\usepackage{graphicx}

\begin{document}

\title{EXCLUSIVE PRODUCTION OF PION PAIRS 
WITH LARGE INVARIANT MASS
IN NUCLEUS-NUCLEUS COLLISIONS
\thanks{This work is
supported by the Polish grant N N202 236640}}

\author{\underline{M.~K{\l}usek-Gawenda} and A.~Szczurek\\ \\
Institute of Nuclear Physics\\
Polish Academy of Sciences\\
PL-31-342 Cracow,\\
Poland\\
E-mail: mariola.klusek@ifj.edu.pl}

\maketitle

\begin{abstract}
\noindent 
The cross section for exclusive $\pi^+\pi^-$ and $\pi^0\pi^0$
meson pairs production in peripheral ultrarelativistic heavy-ion collisions
is calculated at the energy available at the CERN Large Hadron Collider,
i.e., $\sqrt{s_{NN}}=3.5$ TeV. 
The cross section for elementary $\gamma\gamma\to\pi\pi$ process
is calculated with the help of the pQCD Brodsky-Lepage
approach with the distribution amplitude used recently to describe
the pion transition form factor measured by the BaBar Collaboration.
\end{abstract}



\markboth{\large \sl \hspace*{0.25cm}\underline{M.~K{\l}usek-Gawenda} \& A.~Szczurek
\hspace*{0.25cm} Low-$x$ Meeting 2011} {\large \sl \hspace*{0.25cm} TEMPLATE FOR THE
LOW$x$ 2011 MEETING PROCEEDINGS}

\section{Introduction} 
%
\begin{figure}[htb]
\centering
\includegraphics[width=.3\hsize]{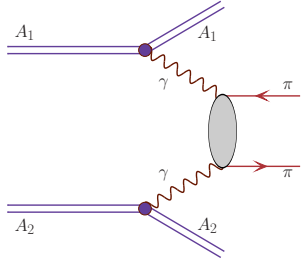}
\vskip -0.3cm
\caption{The Feynman diagram for the formation of the pion pair. 
The $A_1$ and $A_2$ letters denote the $^{208}Pb$ nuclei.}
\end{figure}
It is known that ultrarelativistic colliding heavy ions are a source of high-energy
$\gamma\gamma$ collisions. We present realistic cross section
for exclusive electromagnetic production of two neutral and 
two charged pions in coherent photon-photon processes
in ultrarelativistic heavy-ion collisions.
We consider $Pb Pb \to Pb Pb \pi^{+} \pi^{-}$ and
$Pb Pb \to Pb Pb \pi^{0} \pi^{0}$ reactions.
In Fig.~1 we show the basic mechanism of the exclusive production of 
$\pi^+\pi^-$ and $\pi^0\pi^0$ meson pairs.
To calculate the correct cross section, we have to 
take into account several important factors.
First we include realistic charge densities in nuclei.
The validity of this ingredient has been presented 
in our previous publications where we have studied the production
of $\rho^0\rho^0$ \cite{KS_rho},
   $\mu^+\mu^-$ \cite{KS_muon},
   heavy-quark heavy-antiquark \cite{KS_quark}
as well as
   $D\bar{D}$ \cite{LS2010} pairs.
The next step is a correct description of the elementary
cross section. This was done using the approach proposed by
Brodsky and Lepage. They made a first prediction of the LO pQCD approach
\cite{BL81}. The pQCD amplitude for the $\gamma\gamma\to\pi\pi$
reaction depends on the pion distribution amplitude. 
It was believed for long time that the pion distribution amplitude is close
to the asymptotic form. This turned out to be inconsistent with
recent results of the BaBar Collaboration \cite{BABAR} 
for the pion transition form factor for large photon virtualities.  
%
\section{The elementary cross section for $\gamma\gamma\to\pi\pi$}
%
\begin{figure}[htb]
\centering
\includegraphics[width=.3\hsize]{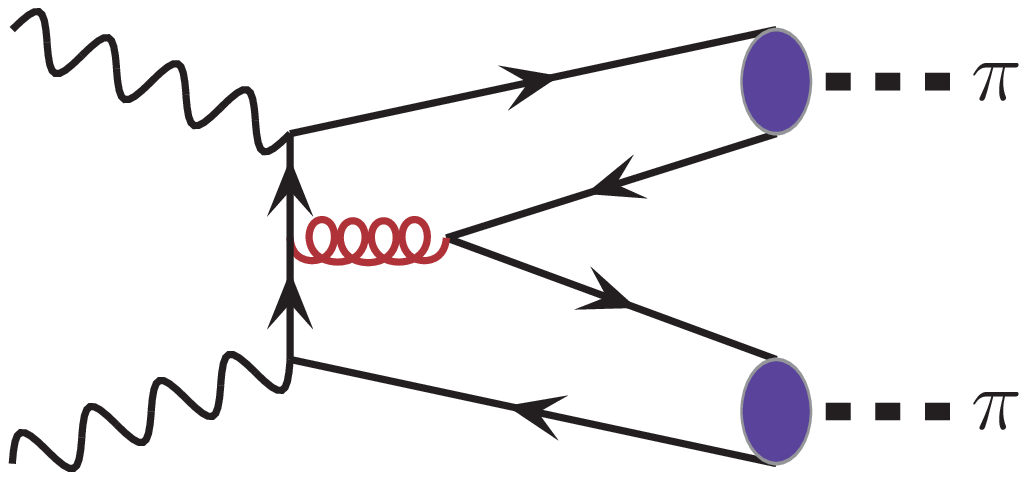}
\includegraphics[width=.3\hsize]{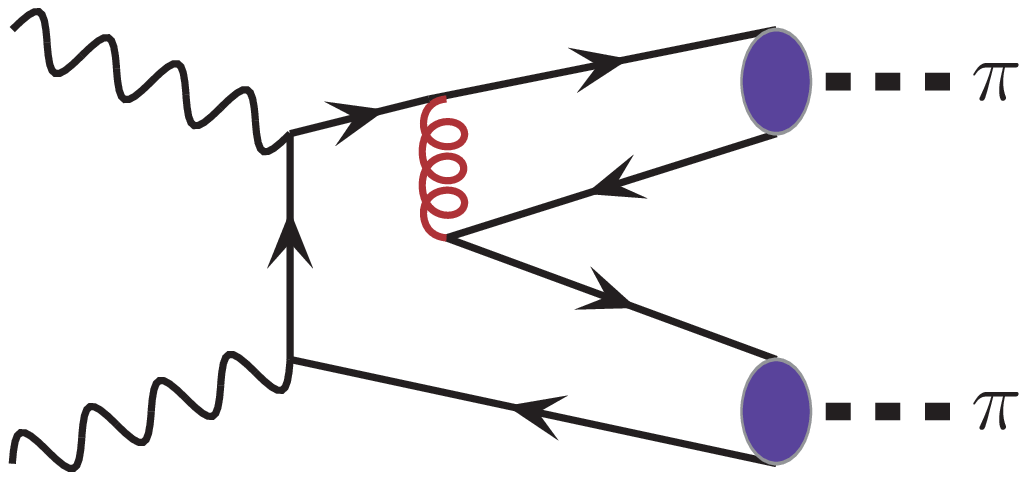}
\includegraphics[width=.3\hsize]{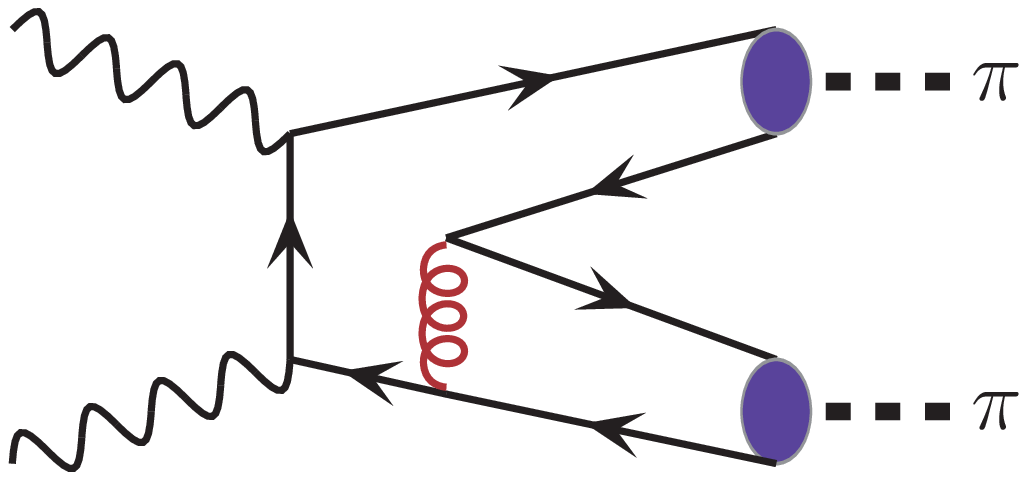}
\includegraphics[width=.3\hsize]{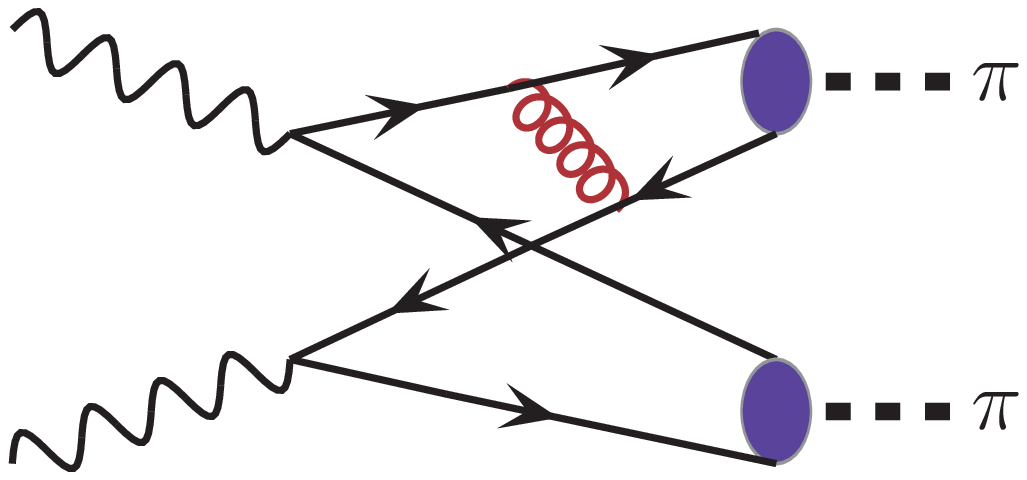}
\includegraphics[width=.3\hsize]{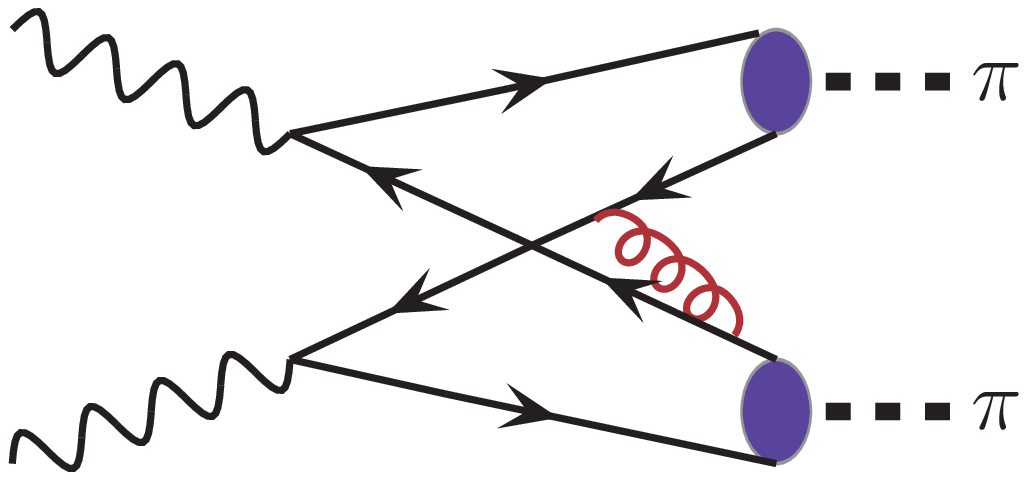}
\vskip -0.3cm
\caption{Feynman diagrams describing the 
$\gamma\gamma  \to \pi\pi$
amplitude in the LO pQCD.}
\end{figure}
Basic diagrams of the Brodsky and Lepage formalism are shown in Fig.~2.
The invariant amplitude for the initial helicities of two photons 
can be written as:
\begin{eqnarray}
 \mathcal{M}  \left( \lambda_1, \lambda_2 \right)  = 
 \int_0^1 dx \int_0^1 dy \, \phi_\pi \left( x, \mu^2_x \right) 
 T^{\lambda_1 \lambda_2}_H \left( x,y, \mu^2 \right) \phi_\pi \left( y, \mu^2_y \right)
 \times F^{pQCD}_{reg} \left(t,u \right),
\end{eqnarray}
where
$\mu_{x_{/y}} = min \left( x_{/y}, 1-x_{/y}  \right) \sqrt{s(1-z^2)}$; 
$z= \cos \theta$ \cite{BL81}.
We take the helicity dependent hard scattering amplitudes from Ref.~\cite{JA}.
These scattering amplitudes are different for $\pi^+ \pi^-$
and $\pi^0 \pi^0$. 
The extra form factor in Eq.~1 was proposed in Ref. \cite{SS2003}:
\begin{equation}
 F^{pQCD}_{reg} \left(t,u \right)= 
 \left[ 1-\exp \left( \frac{t-t_m}{\Lambda_{reg}^2} \right) \right] 
 \left[ 1-\exp \left( \frac{u-u_m}{\Lambda_{reg}^2} \right)  \right],
\end{equation}
where $t_m=u_m$ are the maximal kinematically allowed values of $t$ and $u$. 
This form factor excludes the region of small Mandelstam 
$t$ and $u$ variables which is clearly of nonperturbative nature.
\begin{figure}[htb]
\centering
\includegraphics[width=.45\hsize]{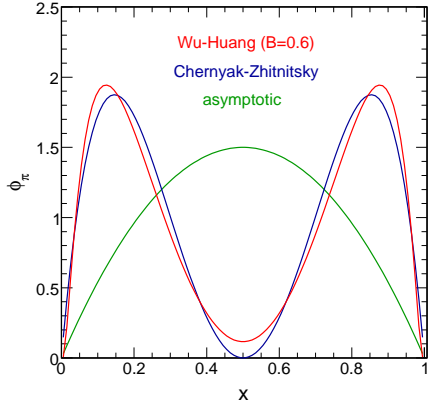}
\vskip -0.3cm
\caption{The quark distribution amplitude of the pion.}
\end{figure}

The distribution amplitudes are subjected to the ERBL pQCD evolution \cite{ER, BL79}.
The scale dependent quark distribution amplitude of the pion \cite{Muller,AB}
can be expanded in term of the Gegenbauer polynomials:
\begin{equation}
\phi_\pi \left( x, \mu^2 \right) = 
\frac{f_\pi}{2 \sqrt{3}}
6x \left( 1-x \right) \sum_{n=0}^{\infty '} C_n^{3/2} 
\left(2x-1 \right) a_n \left(\mu^2 \right)  ,
\end{equation}
where the expansion coefficients $a_n \left(\mu^2 \right)$
depend among others on the form of the distribution amplitude
$\phi_\pi \left( x, \mu_0^2 \right)$.
Different distribution amplitudes have been used in the past 
\cite{BL81,Ch,AB}.
Wu and Huang \cite{WH} proposed recently a new solution:
\begin{eqnarray}
 \phi_\pi \left( x, \mu_0^2 \right) & = 
 & \frac{\sqrt{3}A \, m_q \beta}{2 \sqrt{2} \pi^{3/2} f_\pi} 
 \sqrt{x \left( 1-x \right)} 
 \left( 1+B \times C_2^{3/2} \left(2x-1 \right) \right) \nonumber \\
 &\times& \left( \mbox{Erf} \left[ \sqrt{\frac{m_q^2+\mu_0^2}{8 \beta^2 x \left(1-x \right)}} \right] 
 - \mbox{Erf} \left[ \sqrt{\frac{m_q^2}{8 \beta^2 x \left(1-x \right)}} \right]\right).
 \label{eq.WH}
\end{eqnarray}
This pion distribution amplitude at the initial scale 
($\mu_0^2=1$ GeV$^2$) is controlled by 
the parameter B. It has been found that the BABAR data 
at low and high momentum regions can be well described by setting B to be around 0.6. 
As seen from Fig.~3, this pion distribution amplitude is rather close to the well know 
Chernyak-Zhitnitsky \cite{CZ} distribution amplitude.

Finally, the total (angle integrated) cross section
can be calculated as:
\begin{equation}
\sigma \left(\gamma \gamma \to \pi \pi \right) = 
\int \frac{2 \pi}{4 \cdot 64 \pi^2 W^2 } \frac{p}{q} 
\sum_{\lambda_1, \lambda_2} 
\left|  \mathcal{M}  \left( \lambda_1, \lambda_2 \right) \right|^2 dz \;.
\end{equation}
%
\section{The nuclear cross section for the $PbPb \to PbPb \pi \pi$ process }
%
%
\begin{figure}[htb]
\centering
\includegraphics[width=.45\hsize]{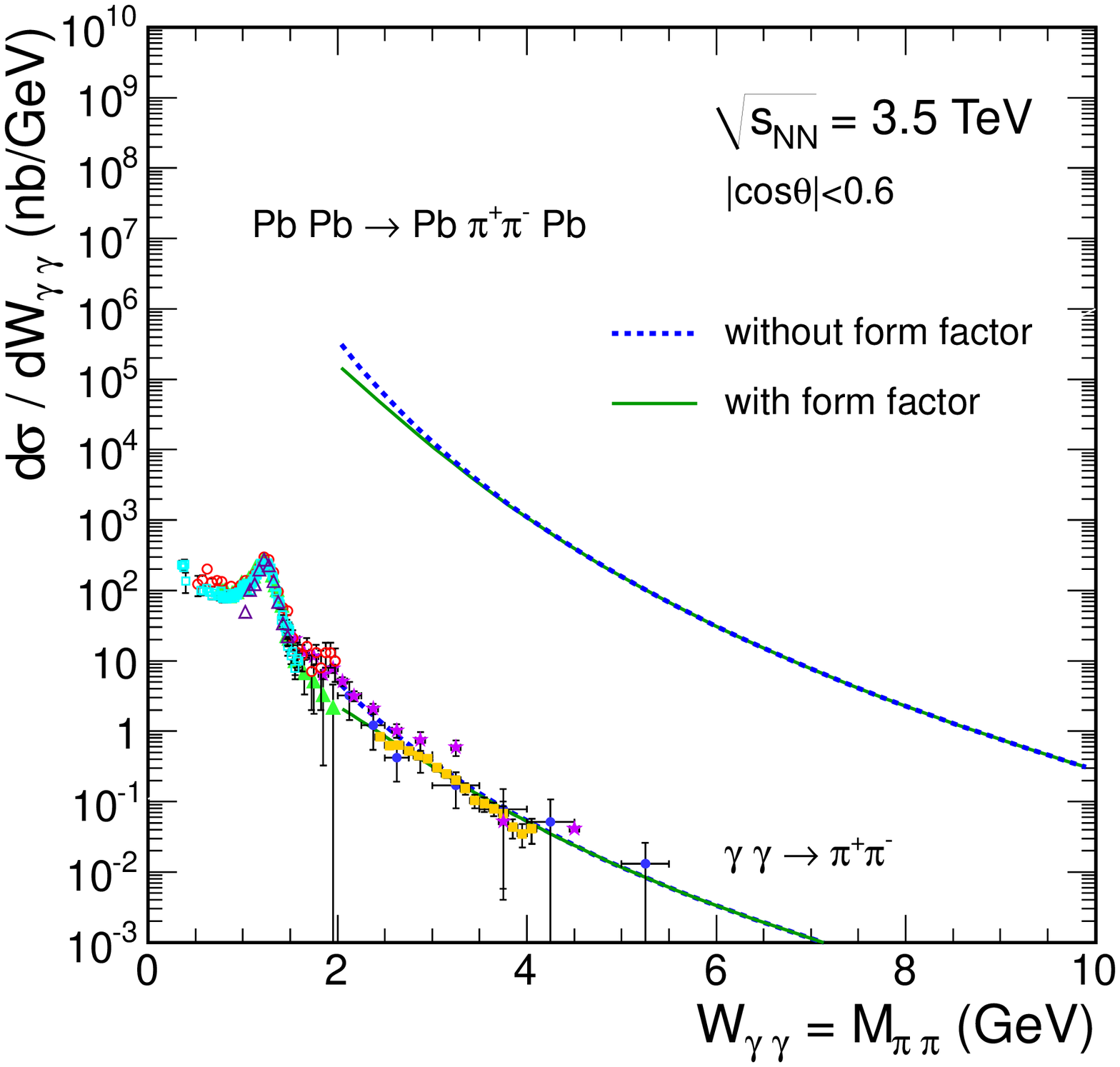}
\includegraphics[width=.45\hsize]{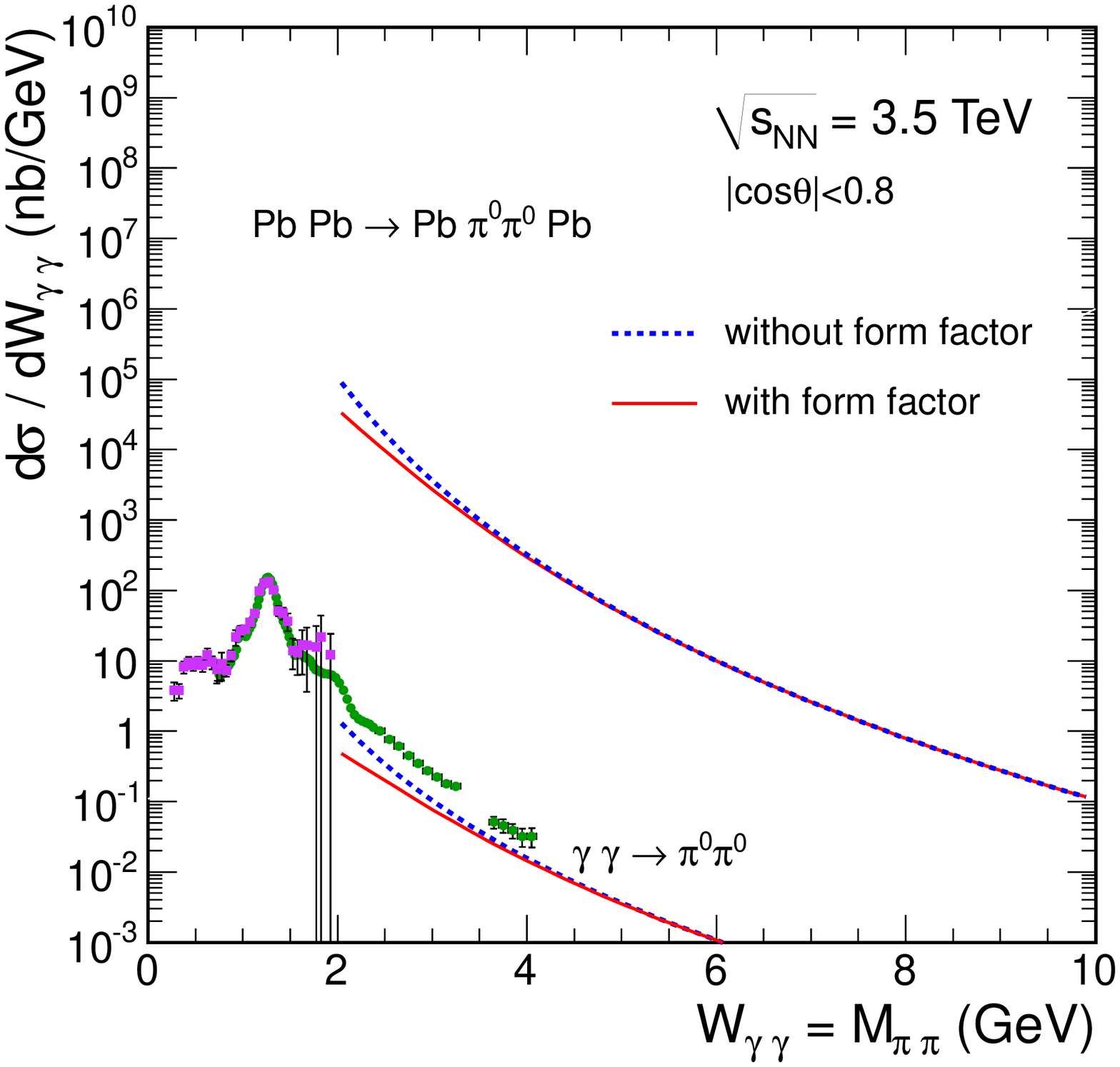}
\vskip -0.3cm
\caption{The nuclear (upper lines) and elementary (lower lines) cross section
   as a function of photon--photon subsystem energy 
   $W_{\gamma \gamma}$ in the b-space EPA.}
\end{figure}
The nuclear cross section has been calculated with the help of b-space
equivalent photon approximation (EPA).
This approach allows to separate peripheral collisions of nuclei
($b>R_1+R_2 \approx14$ fm).
A compact formula for calculating the total cross section takes the form:
 \begin{eqnarray}
\sigma \left(Pb Pb \rightarrow Pb Pb \pi \pi ; W_{\gamma \gamma}\right)
= \int  {\hat \sigma}\left(\gamma\gamma\rightarrow \pi \pi; W_{\gamma \gamma}
\right) \theta \left(|{\bf b}_1-{\bf b}_2|-2R_A \right)& &
\nonumber \\
\times   N \left(\omega_1,{\bf b}_1 \right) N\left(\omega_2,{\bf
b}_2 \right)2 \pi b \, {\rm d} b \, {\rm d} \overline{b}_x \, {\rm
d} \overline{b}_y \frac{W_{\gamma \gamma}}{2} {\rm d}W_{\gamma
\gamma} {\rm d} Y & \, & .
\end{eqnarray}
The details of its derivation can be found in our last papers \cite{KS_rho, KS_muon, KS_quark}.

In Fig.~4 we show distribution in the two-pion invariant mass.
Here we have taken experimental limitations usually used
for the $\pi \pi$ production in $e^+ e^-$ collisions. In the same figure we show our results
for the $\gamma \gamma$ processes extracted from the $e^+ e^-$ collisions
together with the corresponding nuclear cross sections for $\pi^+ \pi^-$ (left panel)
and $\pi^0 \pi^0$ (right panel) production.
We show the results for the case when we include
the extra form factor in Eq.~2 (solid lines) and
for the case when $F^{pQCD}_{reg} \left(t,u \right)=1$ (dashed lines).
One can see that a difference occurs only at
small energies which is not the subject of the present analysis. 
Above $\sqrt{s_{\gamma \gamma}} >$ 3 GeV the two approaches
coincide. By comparison of the elementary and nuclear cross sections
we see a large enhancement of the order of 10$^4$ which is somewhat less
than $Z_1^2 Z_2^2$ one could expect from a naive counting.
%
\section{Outlook}
%
\begin{figure}[htb]
\centering
\includegraphics[width=.45\hsize]{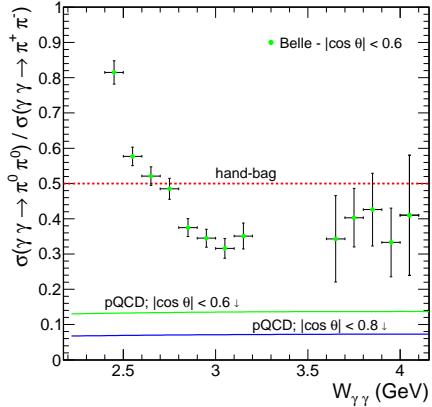}
\vskip -0.3cm
\caption{Ratio of the cross section for the 
   $\gamma \gamma \to \pi^0 \pi^0$ process to that for the
   $\gamma \gamma \to \pi^+ \pi^-$ process.}
\end{figure}
In Fig.~5 we show the ratio of the
cross section for the $\gamma \gamma \to \pi^0 \pi^0$ 
process to that for the $\gamma \gamma \to \pi^+ \pi^-$ process.
The dashed line represents the hand-bag model \cite{HB} result
and the solid lines is for the Brodsky-Lepage pQCD approach. 
For larger $z=\cos \theta$
the ratio is smaller which means that the ratio is $z$ dependent. 
The ratio is practically independent of the collision energy.
In the present calculations
the $z$-averaged ratio for $|\cos \theta|<$ 0.6 is about $0.12$.
The experimental data points are in between the predictions of the BL pQCD
approach and the hand-bag model which further clouds the situation.
More results one can find in our last paper \cite{KGSz2011}.

From Fig.~5 we can conclude 
that other mechanisms are necessary
to correctly describe the elementary cross section.
We are in the process of including
pion exhange, resonances and high-energy $\pi\pi$ rescatterings.

\end{document}